\documentclass[10pt,letterpaper,onecolumn]{article}
\usepackage{amsmath,amssymb,graphicx,cite,url,float,dsfont}
\usepackage{lettrine}
\usepackage{lineno}
\begin{document}
\title{\textsf{\textbf{Three-dimensional imaging with single-frame jigsaw-puzzle-reorganized sinusoidal fringe using multi-pixel axial flat brush scanning}}}

\author{Wen-Kai Yu$^{1,2}$}
\footnotetext[1]{Center for Quantum Technology Research, School of Physics, Beijing Institute of Technology, Beijing 100081, China. Correspondence and requests for materials should be addressed to W.-K.Y. (email: yuwenkai@bit.edu.cn)}
\footnotetext[2]{Key Laboratory of Advanced Optoelectronic Quantum Architecture and Measurements of Ministry of Education, School of Physics, Beijing Institute of Technology, Beijing 100081, China}

\date{}

\maketitle

\renewenvironment{abstract}{%
    \setlength{\parindent}{0in}%
    \setlength{\parskip}{0in}%
    \bfseries%
    }{\par\vspace{-6pt}}

\begin{abstract}
Structured-light three-dimensional (3D) imaging can achieve 3D shape of a stationary object via one or more pixelated array cameras with phase-shifting illumination. In order to extend 3D imaging to moving scenarios, we propose a 3D imaging method with double projection of a single-frame modulated light pattern and a sampling pattern. It can continuously image the moving 3D scene by making multi-pixel detector axial flat brush scan along the motion axis. Utilizing spatial multiplexing for multiple single-pixel imaging, each single-pixel does not need to keep staring at some part of the object, avoiding motion blur problem. The performance of our method has been demonstrated by numerical simulations. Given this, we believe that the technique paves the way to practical applications including product line 3D monitoring.
\end{abstract}

\section*{\textsf{Introduction}}
\lettrine[lines=2]{T}{ree-dimensional} (3D) imaging based on structured illumination reconstructs the surface shape of a target from the geometric distortion (displacement) of the pattern that appears when one or several pixelated array cameras see the target from different perspectives than that of the projected light \cite{Cai2017}. Although measuring a point repeatedly can enhance the quality of the depth image, a better way is to repeat the projected gradient as stripes so that the adjacent parallel lines have different brightness. The widely used strategy is four-step phase-shift sinusoid illumination \cite{Zhang2015}, whose four patterns of stripes are generated by projecting light through a spatial light modulator (SLM) \cite{Jiang2017}. Structured light 3D scanning is one of famous 3D imaging technologies, which sweeps the line light across the scene, and uses two line array detectors to track the whole line. Most 3D scanning methods are based on triangulation, which uses the basic trigonometric function to recover the images. Another traditional method optical-mechanically scans a laser spot across the entire scene. However, regardless of applying array detection, line sweep, or point sweep, they all rely on point-by-point correspondence between the target and the image plane, and fail to exploit the spatial relevance between pixels of the target, resulting in a low signal-to-noise ratio (SNR).

In recent years, a large number of single-pixel imaging (SPI) techniques \cite{Duran2012,Bian2016,MJSunNC2016,Phillips2017} have sprung up to solve aforementioned problems, by collecting the total intensity into a point detector, and providing advantages for applications where the array detectors are unavailable \cite{Radwell2014,Liu2017}. Most of these SPI approaches are computational imaging schemes, usually referring to ghost imaging (GI) \cite{ShihPRL1995,Boyd2002,Shapiro2008,Zhao2012,YuOE2014,BSunSci2013} and compressive imaging \cite{Baraniuk2008,Studer2012,YuSR2014,YuAO2015,YuOC2016}. GI, as a statistical mechanism, retrieves the image from the second-order intensity correlation between the modulated patterns and the bucket/single-pixel signal. There are many GI methods aiming at increasing the SNR, such as differential ghost imaging (DGI) \cite{Ferri2010} and correspondence imaging \cite{Luo2011,Luo2012,YuCPB2015}. To our knowledge, they all require a larger number of measurements than the dimensions of the object image. Fortunately, compressed sensing (CS) \cite{Donoho2006,Candes2006} has been presented to offer a way to greatly reduce the sampling ratio, as long as the object image is sparse or compressive. But its convex optimization procedure for solving ill-posed image recovery problem often leads to a huge computational overhead and a non-deterministic inaccurate solution \cite{YuOC2017,Sun2018}. Actually, all these SPI techniques exchange the acquisition time and the computational complexity for spatial pixel resolution \cite{Edgar2019}. This problem has been the main bottleneck for SPI being used in practical real-time applications \cite{YuAOtrack2015,Wang2017}, especially for 3D imaging of relative moving targets (e.g., imaging 3D objects on a conveyor belt, or earth observation and navigation on unmanned aerial systems \cite{Yu2018}). The cause of this problem is due to the fact that SPI is based on a typical staring imaging mechanism and there exists a lot of spatial redundancy in motion detection. To address this problem, a push-broom scanning strategy \cite{Ma2019}, derived from the row-scanning method \cite{Yu2014,YuAO2015,YuOC2016} of the digital micromirror device (DMD), has been proposed and used in a two-dimensional (2D) laser radar system, but with a very large augmented diagonal measurement matrix, which greatly limits its cross dimension and image quality.

In this work, we propose a 3D imaging method by projecting both single-frame jigsaw-puzzle-reorganized sinusoidal pattern and sampling pattern, then making multiple detection pixels axial flat brush scan along the relative movement direction. Note that we make the line array detector be parallel to the motion axis rather than perpendicular/across to the axis. By this means, it takes full advantage of spatial multiplexing and actually performs multiple parallel SPI. As a result, our method does not need to make single-pixel be sequentially staring at some part of the targets, getting rid of motion blur problem. The simulation results has shown its potential of 3D imaging of the relative moving targets and its reconstruction ability of stationary 3D scene of long range. Therefore, this technology can be widely used in 3D reproduction of the museum artifacts, 3D digitization of the human body, oral 3D scan, virtual vision, autonomous driving, and so on.

\section*{\textsf{Principles and Methods}}
\begin{figure}[htbp]
\centering\includegraphics[width=0.95\linewidth]{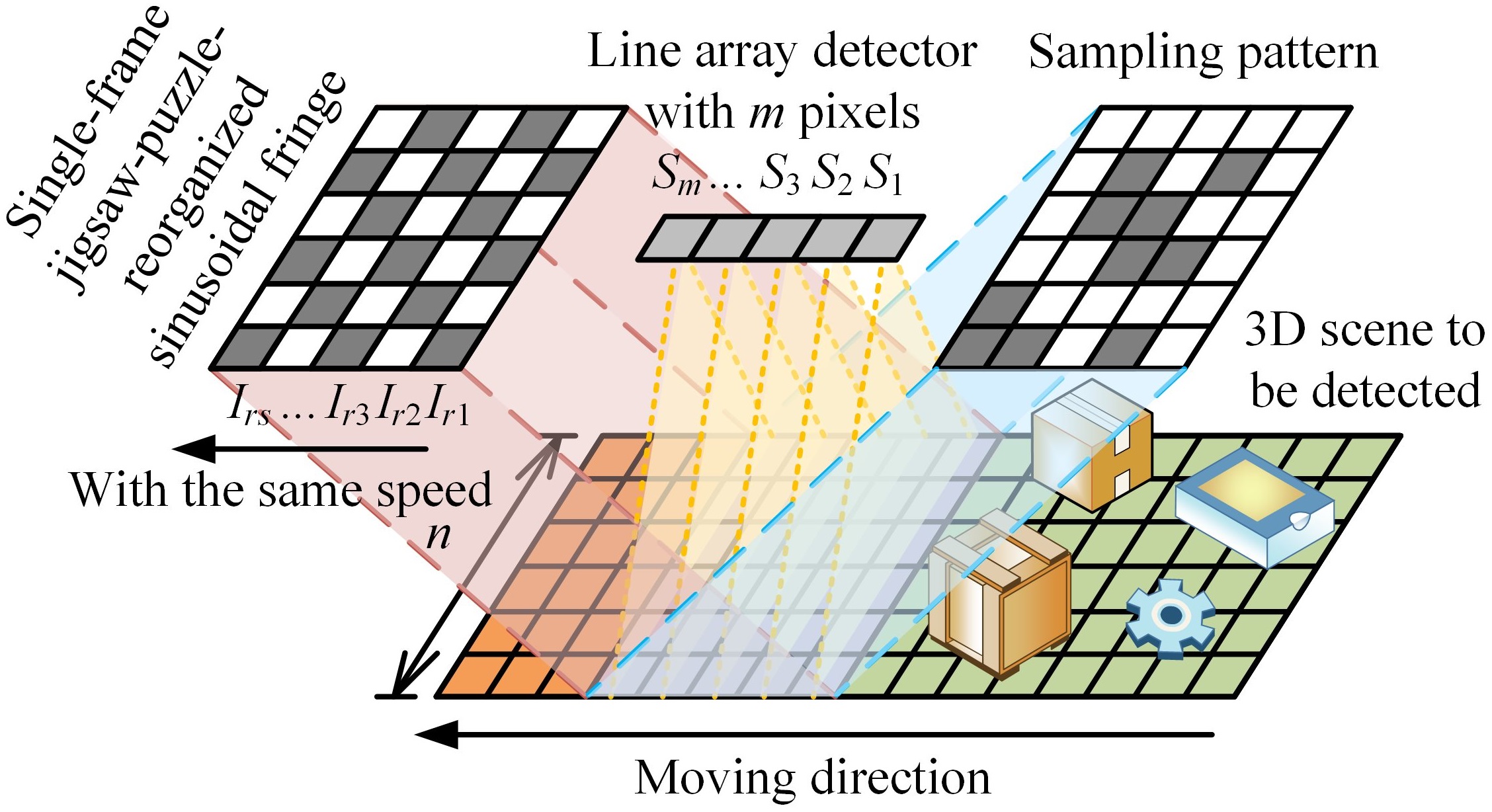}
\caption{Schematic of 3D imaging with single-frame jigsaw-puzzle-reorganized sinusoidal fringe using multi-pixel axial flat brush scanning. The first projector on the left can be either fixed and encoded with a real-time translational fringe or moving with a fixed fringe. The translational velocity of the first fringe pattern (or projector) is the same as that of the conveyor belt. The repetition period of the entire fringe pattern is related to the slope of the sinusoidal fringe. It is best to apply the end-to-end approach, just like a rotating lamppost in front of the barber shop. The second projector projects a deterministic sampling pattern on a fixed region, with the size less than or equal to that of the first projected sinusoidal fringe. It is worth mentioning that if we use time-varying pattern (i.e., the pattern changes for every moving steps) on the second projector, this pattern should be random, which will results in measurement redundancy \cite{Yu2019}. For the time-invariant pattern case, the above process is equivalent to the sampling pattern being point-multiplied by a jigsaw-puzzle-reorganized strip pattern that keeps moving down in each step. The line array detector with $m$ pixels will sample the projected area of the scene.}
\label{fig:Schematic}
\end{figure}

Generally speaking, the targets to be detected cannot be stationary all the time, i.e., there exists a relative movement between the target scene and the optical imaging system. Now, we assume the detection system to be stationary, while the 3D scene moves at a constant speed in a certain direction. The schematic of our proposed 3D method is given in Fig.~\ref{fig:Schematic}.

The 3D scene is set on a moving platform, such as a conveyor belt. One may easily imagine that baggage screening in airports, subway stations, railway stations or product inspection on the assembly line are all suitable for this situation. The principle is the same for the moving detection systems, just like in the field of navigation and landing of unmanned aerial vehicles. In the illustrated 3D imaging system (see Fig.~\ref{fig:Schematic}), it projects a single-frame calibrated rectangular pattern and a sampling pattern onto the moving platform, the latter covers a relatively fixed projection area of the platform (represented by the purple region) while the former covers a larger region (it is not necessary to be fixed). The patterns used will be detailed later. Here we place a line array detector above the detected platform, with $m$ pixels parallel (rather than perpendicular) to the motion axis. The imaging scene can be split into a lot of columns, then the aforesaid $m$ pixels collects $m$ column-total-intensities separately with help of a cylindrical lens, that is, one pixel records the total intensity of one column of the scene. When the 3D scene to be sampled starts to move, $m$ pixels on the line array detector will apply $m$ column-strips (constituting a single-frame pattern) to measure each column signal in the scene at the same time, taking full advantage of spatial multiplexing, i.e., one column signal being sampled for $m$ times. This process is like treating multiple pixels as a flat brush and brushing it over the entire detected 3D target along the opposite motion direction of the scene. Usually, it is not difficult to control the moving speed precisely, even for unmanned aerial vehicles. Thereby this scheme can be implemented easily.

The principle behind our 3D imaging technique is based on the theorem of structured illumination. Let's first briefly introduce the traditional four-step phase-shifting scheme: it projects four fringe patterns upon the scene, and captures distorted fringe image due to the 3D geometry of the scene, then unwraps the phase map between the stair phase images. There are three main merits of applying phase-shifting technology, i.e., 1) the ability of obtaining pixel-by-pixel spatial resolution, 2) the sensitivity to local surface reflectivity variations, and 3) the robustness against the ambient light. The intensity of a sinusoidal fringe pattern projected onto the target can be mathematically formulated as
\begin{align}
I_i\left(x,y;f_u,f_v\right)=&a(x,y)+b(x,y)\cdot\\\nonumber
&\cos\left(2\pi f_ux+2\pi f_vy+\phi+i\pi/2\right),
\end{align}
where $i=0,1,2,3$, $x$ and $y$ are the coordinates of the image pixels, $f_u$ and $f_v$ denote spatial frequencies of fringes, $a(x,y)$ is the background/average intensity (also known as the direct current (DC) component) and $b(x,y)$ is the modulation intensity or the envelope, $\phi$ stands for the initial phase. The recorded radiance intensity $F_i$ is in proportion to the projected intensity $I_i$ and the surface reflectivity $R$, and can be written as
\begin{align}
F_i\left(x,y;f_u,f_v\right)=&R(x,y)\cdot I_i\left(x,y;u,v\right)\\\nonumber
=&R(x,y)\cdot a(x,y)+R(x,y)\cdot b(x,y)\cdot\\\nonumber
&\cos\left(2\pi f_ux+2\pi f_vy+\phi+\phi(x,y)+i\pi/2\right),
\end{align}
where $R(x,y)\cdot a(x,y)$ and $R(x,y)\cdot b(x,y)$ can be regarded as the modulated background intensity (or environmental illumination noise) and modulated reflectivity intensity, respectively. $\phi(x,y)$ represents the phase to be extracted, which is determined by the scene depth, ranging from 0 to $2\pi$. Set $\phi'(x,y)=2\pi f_ux+2\pi f_vy+\phi+\phi(x,y)$ and $\phi_{o}(x,y)=2\pi f_ux+2\pi f_vy+\phi$, and they can be computed as follows:
\begin{align}
\phi'(x,y)&=\arctan\left(\frac{F_3-F_1}{F_0-F_2}\right),\\
\phi_{o}(x,y)&=\arctan\left(\frac{I_3-I_1}{I_0-I_2}\right).
\end{align}
According to the arctangent function, the achieved phase value ranges from $-\pi$ to $\pi$ with $2\pi$ discontinuities. By using a spatial or temporal phase unwrapping algorithm, which finds the discontinuities locations and compensates the $2\pi$ jumps, we can unwrap the phase and shift it to a range of 0 to $2\pi$ to obtain a continuous phase angle map. After separately performing two-dimensional phase unwrapping on $\phi'(x,y)$ and $\phi_{o}(x,y)$ to obtain two phase maps, we will extract the phase information of the scene via $\phi(x,y)=\phi'(x,y)-\phi_{o}(x,y)$. From above equations, it can be seen that the DC item $R(x,y)\cdot a(x,y)$ is almost completely eliminated, thus the phase recovery is robust to noise. After that, 3D surface can be recovered visually by converting the phase angle map to the depth map or the point cloud map, mainly based on parameters such as detector and projector positions, wavelength, sinusoidal fringe spatial period and the like.

\begin{figure}[htbp]
\centering\includegraphics[width=0.95\linewidth]{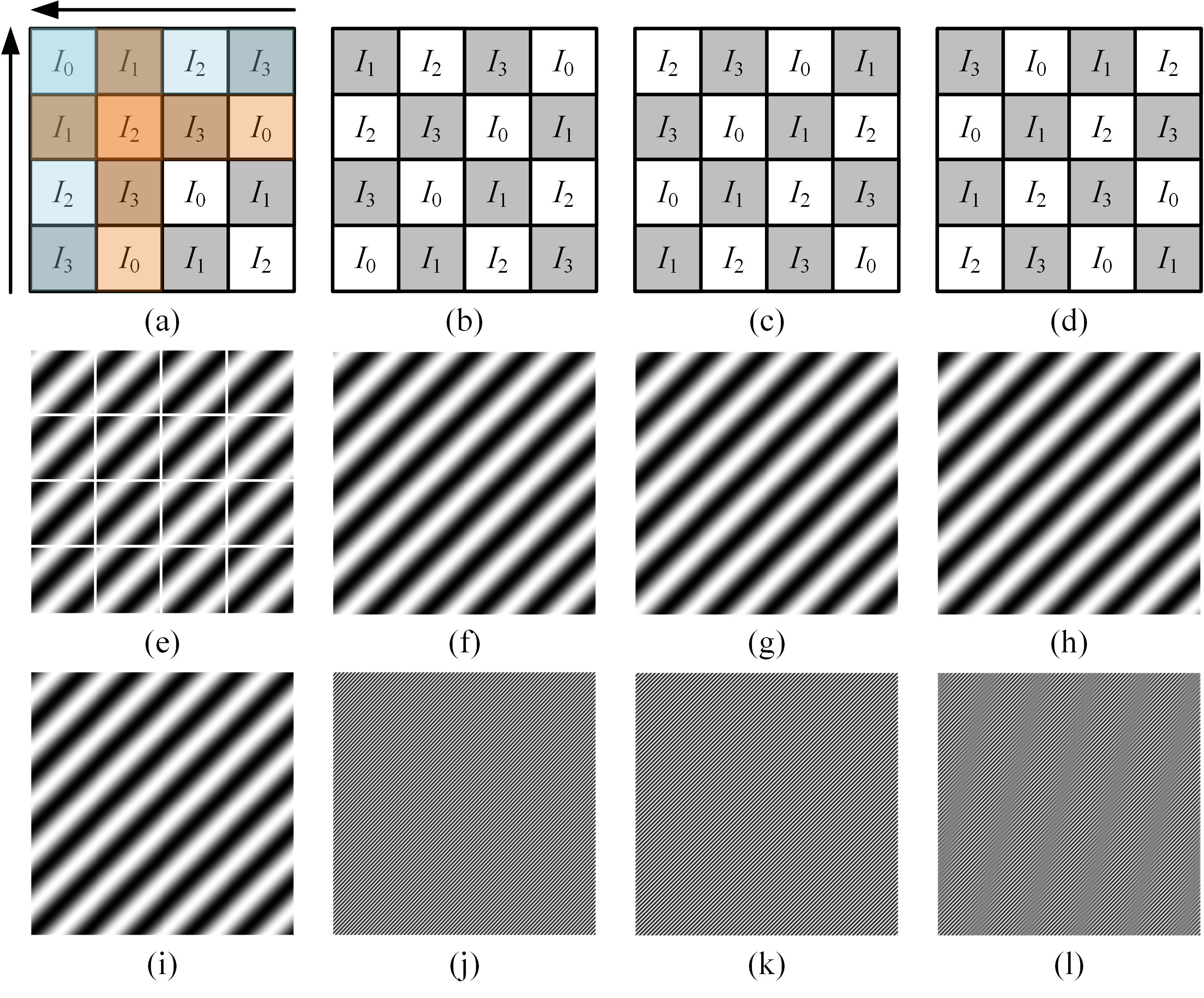}
\caption{Jigsaw-puzzle-reorganized principle and results. (a)--(d) are four jigsaw-puzzle-reorganized matrices. (e) is the reorganized result of $I_0$, $I_1$, $I_2$ and $I_3$, each of which is $62\times62$ pixels, following the arrangement shown in (a), and (f) is the result ($248\times248$ pixels) of removing its white edges. (j) is the reorganized result of four different initial patterns, all of $1\times1$ pixel, the size of new fringe pattern is $256\times256$ pixels. (k) is obtained with the same $I_0$, $I_1$, $I_2$ and $I_3$. (l) is the difference map between (j) and (k).}
\label{fig:JigsawPuzzle}
\end{figure}

In the scenarios of moving objects, the scene passes through the detection area at a constant speed and cannot be turned back. However, if we still want to use traditional four-step phase-shifting illumination, it requires us to repeat the movement of the targets four times in the common sense, which is not allowed in the practical situations. By focusing on the four patterns, we associate with the principle of red, green, blue (RGB) pixel arrangement of a color charge coupled device (CCD), each pixel unit of which generally has four sub-pixels, one for red, one for blue and two on the diagonal for green. But we cannot directly transplant this idea to our scheme, because this principle will split the continuity of the sinusoidal stripes. This is what we don't want to see. Since the sinusoidal stripe itself has periodicity, we try to reorganize the core patterns. Let's set the four phase-shifting stripe patterns as $I_0$, $I_1$, $I_2$ and $I_3$. By carefully choosing the parameters like the slope, the initial phase, the pixel-size, and so on, we sequentially stitch these four patterns together to form a line, then loop shift the sequence to the left in the next three rows, as illustrated in Fig.~\ref{fig:JigsawPuzzle}(a). Under this arrangement, the patterns on the secondary diagonal are the same. Loop left shifting or loop up shifting of this arrangement matrix (the orange row/column moves to the light blue one) will generate another three equivalent matrices, as shown in Figs.~\ref{fig:JigsawPuzzle}(b)--(d). Here we may assume $\phi=3\pi/2$ or $-\pi/2$, then $\cos(\alpha+3\pi/2)=\sin\alpha$ or $\cos(\alpha-\pi/2)=\sin\alpha$ and let $f_u=f_v=1/50$, and set the pixel-size of $I_0$, $I_1$, $I_2$ and $I_3$ to be $62\times62$. Following the arrangement scheme of Fig.~\ref{fig:JigsawPuzzle}(a), we can reorganize the patterns in a form as shown in Fig.~\ref{fig:JigsawPuzzle}(e), then we will obtain one-frame approximate continuous sinusoidal fringes, as shown in Fig.~\ref{fig:JigsawPuzzle}(f). The remaining three arrangements will produce another three sinusoidal fringes, as illustrated in Figs.~\ref{fig:JigsawPuzzle}(g)--(i). Actually, Figs.~\ref{fig:JigsawPuzzle}(f)--(i) are similar to the patterns $I_0$, $I_1$, $I_2$ and $I_3$, respectively, but with 16 times the size. We find that, if the pixel size of each pattern is reduced to one pixel, the rule still works, as shown in Fig.~\ref{fig:JigsawPuzzle}(j) (corresponding to Fig.~\ref{fig:JigsawPuzzle}(a)). In contrast, we also set each pattern to be the same, which generates a fringe pattern shown in Fig.~\ref{fig:JigsawPuzzle}(k). Although Figs.~\ref{fig:JigsawPuzzle}(j) and (k) looks like each other, but their difference map is given in Fig.~\ref{fig:JigsawPuzzle}(l). Actually, if the initial patterns $I_0$, $I_1$, $I_2$ and $I_3$ are the same, their reorganized stripes cannot be used for four-step phase-shifting. In the following, we will adopt the arrangement shown in Fig.~\ref{fig:JigsawPuzzle}(j) to acquire the phase information of the scene. It should be noted that the reconstruction results are the same regardless of using any of the arrangement schemes shown in Figs.~\ref{fig:JigsawPuzzle}(a)--(d). Similar results can be obtained by initially arranging $I_0$, $I_1$, $I_2$ and $I_3$ in reverse order.

As for the intensity part of the 3D scene, we can use multiple single-pixel sampling for reconstruction. Although we use a line array detector here, each pixel of which is actually staring at a particular strip region in the absolute space. By utilizing spatial multiplexing, multiple SPI can be performed simultaneously. In SPI, the spatial resolution is transferred from the detector to the modulated patterns, thus only a single-pixel is enough for 2D image recovery. The resolution of each pattern typically equals to that of the target image $\omega\in\mathds{R}^{N}$. There are many SPI methods, like GI, Hadamard SPI and CS. Since the number of measurements in GI is generally much larger than dimensions of the target, we does not use it here. Hadamard SPI is a technology that applies complete deterministic orthogonal bases with entries $\pm1$ to perform full sampling, i.e., the number of measurements equals to the pixel dimension of the object to be detected. It is known that the naturally ordered Hadamard matrix $H$ of order $2^k=N\geq2$ is a symmetric square matrix, defined as
\begin{equation}
H_{2^k}=\left[{\begin{array}{*{20}{c}}
H_{2^{k-1}}&H_{2^{k-1}}\\
H_{2^{k-1}}&-H_{2^{k-1}}
\end{array}}\right]=H_2\otimes H_{2^{k-1}},
\end{equation}
where $H_1=[1]$, $H_2=\left[{\begin{array}{*{20}{c}}
{1}&{1}\\
{1}&{-1}
\end{array}}\right]$, $\otimes$ represents the Kronecker product. We will have $H^T=H$ and $H^{-1}=\frac{1}{N}H$. Therefore, the measurement process can be written as $c=H\omega+e$, where $c\in\mathds{R}^{N\times1}$ stands for the single-pixel measured vector and $e$ denotes the stochastic noise, and it is convenient for us to fast recover the signal only by matrix multiplication $\omega=\frac{1}{N}Hc$. If the signal to be sampled is inherently sparse or compressive in some invertible (e.g. orthogonal) basis $\Psi$, then we can use CS instead, but with a number of measurements $M=O(K\cdot\log(N/K))$ much fewer than $N$, where $K$ denotes the sparsity of the signal, breaking the Nyquist criterion. Thereby, the pixel dimension of the measurement matrix $A$ is $M\times N$ and then the measured signal becomes $c=A\omega+e$. Since the number of equations is much less than the number of unknowns, the solution to this problem is ill-posed. Fortunately, CS provides many convex optimization (norm-1 or norm-2) algorithms to solve such a linear problem, such as total variation minimization (TVAL3 solver) \cite{CBLi2010} which is used here. In CS, the measurement matrix usually adopts a random matrix, or is generated by randomly disrupting the Hadamard matrix and then taking arbitrary subsets consisting of its $M$ rows. We find that random patterns fail to explore internal relations between the patterns and the image reconstruction, and will cause redundant and blind measurements. By using a cake-cutting sort of Hadamard basis \cite{Yu2019}, we can minimize sampling redundancy and greatly reduce computational overhead. In our scheme, each row of $H$ or $A$ can be transposed to a column pattern ($h_j$ or $a_j$) of $N\times 1$ pixels projected onto the moving scene, covering a stripe region of the scene. As we know, although CS in principle can provide a benefit in reducing acquisition time (i.e., the number of measurements) by leveraging the sparsity of the object, CS is not omnipotent. Especially in the cases of the object signal is not sparse or compressive and an accurate reconstruction is desired, Hadamard SPI will be more suitable. Therefore, a rational choice should be made to acquire a better 3D shape of the target.

Refer to Eq.~(2), $F_0$, $F_1$, $F_2$ and $F_3$ are now pieced together to form a single-frame image $F$, each column of which can be regarded as $i$th strip of the scene i.e., $\omega_i\in\mathds{R}^{N\times1}$. As the scene relatively moves at a certain speed, each pixel of the line array detector will sense sequentially the total intensity of the same stripe of the scene $\omega_i$ but with different column patterns: $c_i=H\omega_i$ or $c_i=A_i\omega_i$. In order to facilitate implementation in the actual systems and to avoid measurement redundancy caused by random patterns, here we use a time-invariant deterministic illuminating pattern for sampling, which means that all $A_i$ or $H_i$ are the same. Unlike the push-broom scanning method \cite{Ma2019}, we will not combine all the linear equations together to form super large augmented equations $C=AO$, where $A=\textrm{diag}\{A_1,A_2,\cdots,A_t\}$, $\Omega=[\omega_1,\omega_2,\cdots,\omega_t]^T$, $C=[c_1,c_2,\cdots,c_t]^T$, diag denotes the diagonal matrix, $T$ represents the transpose symbol. This augmented combination will increase the burden of computation and storage, greatly limiting the pixel-dimension (or spatial resolution) of the reconstructed signal and causing periodic degradation of the reconstructed images. To get rid of this problem, we use distributed parallel computing to recover all $\omega_i$, which will be then stitched into $F$. According to the matrix shown in Fig.~\ref{fig:JigsawPuzzle}(a), the system may either use $F_{col}=\left[{\begin{array}{*{20}{c}}
F_0(i,j)&F_1(i,j+1)&F_2(i,j+2)&F_3(i,j+3)\\
F_1(i,j)&F_2(i,j+1)&F_3(i,j+2)&F_0(i,j+3)\\
F_2(i,j)&F_3(i,j+1)&F_0(i,j+2)&F_1(i,j+3)\\
F_3(i,j)&F_0(i,j+1)&F_1(i,j+2)&F_2(i,j+3)
\end{array}}\right]$ or $F_{row}=F_{col}^T$ to form the jigsaw-puzzle-reorganized sinusoidal fringe, denoted as mode-col and mode-row, respectively. From the top perspective facing the motion direction, mode-col (or mode-row) means that the four phase-shifting modulated values reflected from the same space coordinate $(x,y)$ of the scene are listed in one column (or row) of each $4\times4$ cell, which is like the horizontal (or vertical) axis coordinates of the scene are stretched fourfold. In the actual application scenarios, we can precisely control the speed of the conveyor belt, mode-col (or mode-row) corresponds to forward moving $1/4$ pixel (or 1 pixel) per measurement. According to this positional relationship, we can easily find every values of $F_0(i,j)$, $F_1(i,j)$, $F_2(i,j)$ and $F_3(i,j)$ corresponding to all coordinates. By computing Eqs.~(3)--(4), we can acquire the phase angle map $\phi(x,y)$. Substituting $\phi(x,y)$ into Eq.~(2) gives the surface reflectivity $R(x,y)$. Generally, the item $a(x,y)$ in Eq.~(2) is constant and can be easily measured experimentally. For simplicity, we set $a(x,y)$ to 0, then $R(x,y)\cdot a(x,y)$ is also 0. So directly divide $F$ by the jigsaw-puzzle-reorganized reflected $\cos$ item (the one after being modulated) generates $R(x,y)$. Through the above steps, we can obtain the intensity (surface reflectance) and depth information (phase angle) of the moving scene at the same time.

\section*{\textsf{Simulations and Results}}
\begin{figure}[htbp]
\centering\includegraphics[width=0.95\linewidth]{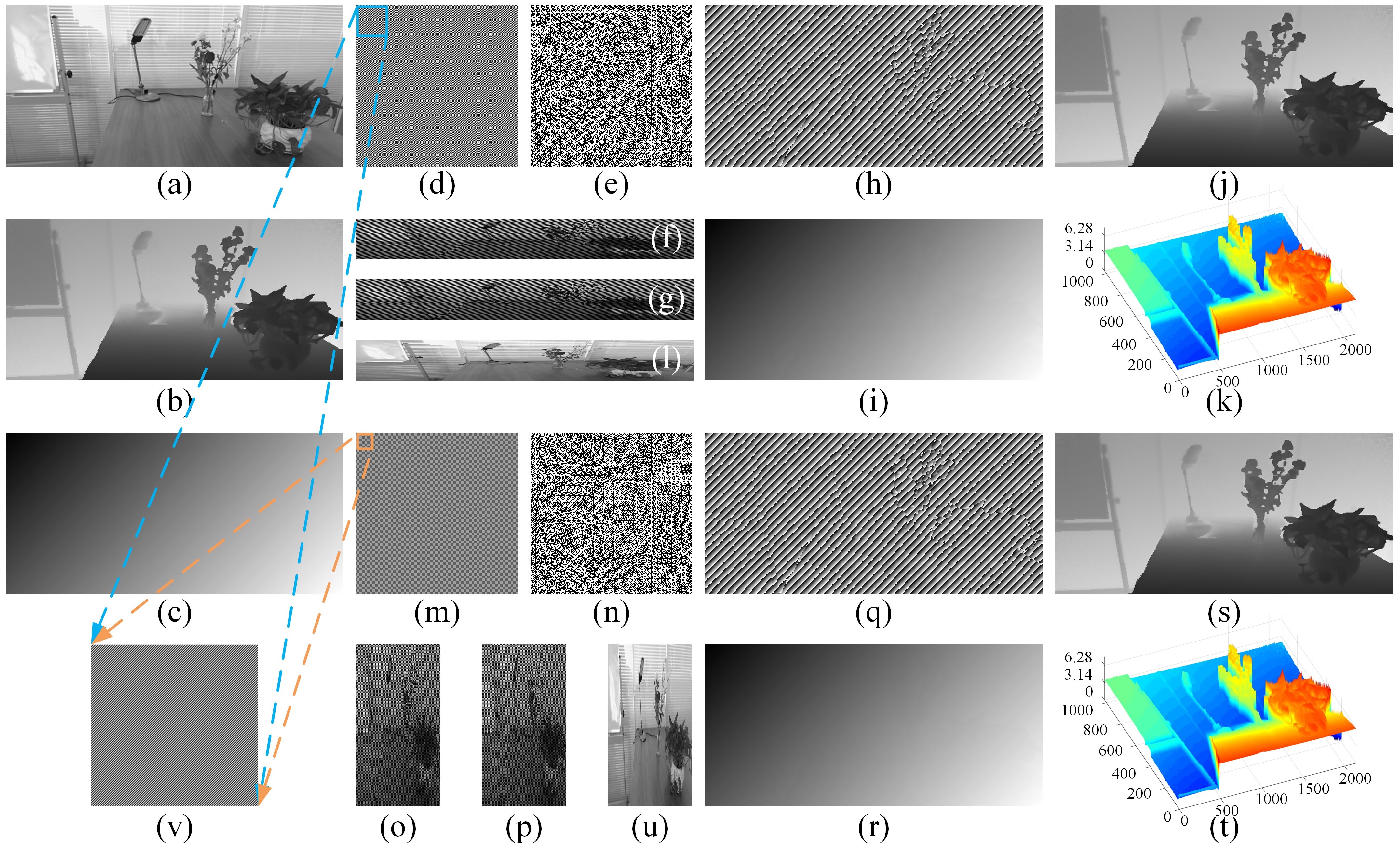}
\caption{Simulation process using Hadamard full sampling for mode-col mode-row. (a) The original surface reflectance image of $1024\times2140$ pixels. (b) The original unwrapped phase angle $\phi(x,y)$, computing from $I_1$, $I_2$, $I_3$ and $I_4$. (c) The unwrapped phase angle image of $\phi_o(x,y)$. (d) The jigsaw-puzzle-reorganized sinusoidal fringe of $1024\times1024$ pixels. (e) The Hadamard matrix of order 1024. (f) The scene is projected with the jigsaw-puzzle-reorganized sinusoidal fringe by the first projector on the left. For mode-col, the horizontal axis coordinates of the scene are stretched four times. (g) After performing Hadamard full sampling, we get the reconstructed image $F$. (h) The wrapped phase image $\phi'(x,y)$ is extracted from $F_{col}$. (i) Unwrapping the phase to estimate the actual phase angle from the wrapped function $\phi'(x,y)$. (j) The differential image (also the phase angle $\phi(x,y)$ of the scene) between (i) and (c). (k) is the 3D display of (j). (l) The surface reflectivity $R(x,y)$ is retrieved by substituting $\phi(x,y)$ into Eq.~(2). Similarly, (m)--(u) give the reconstruction process for mode-row. (m) The jigsaw-puzzle-reorganized sinusoidal fringe of $4096\times4096$ pixels. (n) The Hadamard matrix of order 4096. (o) is the reflected intensity image of the scene after projection with the vertical axis coordinates being stretched fourfold. (p) The reconstructed image $F$ of mode-row. (q) The wrapped phase image $\phi'(x,y)$. (r) Unwrapped phase image. (s) The phase angle map obtained by subtracting (c) from (r). (t) The 3D display of (s). (u) The recovered surface reflectivity $R(x,y)$ of mode-row. (v) is enlarged image of (d) and (m).}
\label{fig:Simulation}
\end{figure}

In order to test the reconstruction performance of our method, some numerical simulations are performed. Fig.~\ref{fig:Simulation} presents our simulation process. We took a picture of our office with Microsoft Kinect 2.0 (see Fig.~\ref{fig:Simulation}(a), the size is $1024\times2140$ pixels) and successfully converted its point cloud data into a phase angle map with a value range from 0 to $2\pi$, as shown in Fig.~\ref{fig:Simulation}(b). The unwrapped phase angle image of $\phi_o(x,y)$ is given in Fig.~\ref{fig:Simulation}(c). For this demonstration, we use the Hadamard full sampling method. As mentioned before, there are two modes: mode-col and mode-row, corresponding to four times stretching on the horizontal axis and the longitudinal axis, respectively. As a sequence, we generate a jigsaw-puzzle-reorganized sinusoidal fringe of $1024\times1024$ pixels together with a Hadamard pattern of the same size for mode-col, and generate a sinusoidal fringe of $4096\times4096$ pixels and a Hadamard pattern with the same size for mode-row, as shown in Figs.~\ref{fig:Simulation}(d)--(e) and \ref{fig:Simulation}(m)--(n). These two pairs of patterns are projected onto the moving scene in mode-col and in mode-row, respectively. Since the sinusoidal stripe pattern moves in the same direction as the scene with the same constant speed, the jigsaw-puzzle-reorganized sinusoidal fringe illuminates the scene relatively statically, as shown in Figs.~\ref{fig:Simulation}(f) and \ref{fig:Simulation}(o). After performing Hadamard full sampling, we can reconstruct the reflected images (see Figs.~\ref{fig:Simulation}(g) and \ref{fig:Simulation}(p)) via matrix multiplication. Then according to the matrix configuration of $F_{col}$ and $F_{row}$, we will extract the wrapped phase images, as shown in Figs.~\ref{fig:Simulation}(h) and \ref{fig:Simulation}(q), which will be turned into unwrapped phase maps, see Figs.~\ref{fig:Simulation}(i) and \ref{fig:Simulation}(r). By subtracting Fig.~\ref{fig:Simulation}(c) from Figs.~\ref{fig:Simulation}(i) or \ref{fig:Simulation}(r), we will obtain the phase angle $\phi(x,y)$ as well as its 3D display of the scene, as presented in Figs.~\ref{fig:Simulation}(j)--(k) and \ref{fig:Simulation}(s)--(t). By substituting $\phi(x,y)$ into Eq.~(2), we can easily compute the surface reflectivity $R(x,y)$, as shown in Figs.~\ref{fig:Simulation}(l) and \ref{fig:Simulation}(u). It is worth mentioning that in the step of getting the unwrapped phase images, there will be many strips of $2j\pi$ jumps at the presence of measurement noise, which can be compensated by finding their positions and subtracting the extra $2j\pi$ part that is beyond the angle range. The simulation results have demonstrated that the reconstructions of our method are perfect with Hadamard full sampling in both mode-col and mode-row.

\begin{figure}[htbp]
\centering\includegraphics[width=0.95\linewidth]{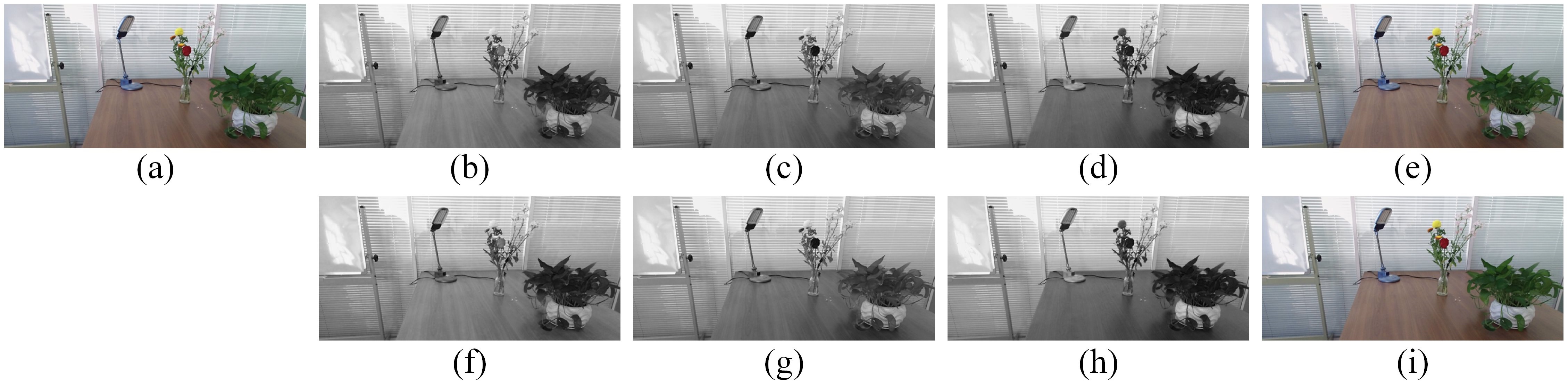}
\caption{Color reconstructions for mode-col mode-row. (a) The original scene of $1024\times2140$ pixels. (b)--(d) and (f)--(h) are retrieved $R(x,y)$ of the RGB channels of mode-col and mode-row, respectively. It should be noted that these images are resized to $1024\times2140$ pixels, by computing the average of each columns (or rows) in each cell, corresponding to mode-col and mode-row. (e) and (i) are the color image obtained by synthesizing the three components (b)--(d) and (f)--(h), respectively. The results are also acquired via the Hadamard full sampling approach.}
\label{fig:Color}
\end{figure}

We also test our method with a color scene (see Fig.~\ref{fig:Color}(a)). By using mode-col and mode-row, we can get the surface reflectance reconstruction of the RGB channels, see Figs.~\ref{fig:Color}(b)--(d) and Figs.~\ref{fig:Color}(f)--(h), respectively, then separately merge them into a color map, as shown in Figs.~\ref{fig:Color}(e) and \ref{fig:Color}(i). All these are finished with Hadamard full sampling, while the phase angle map of the color scene recovered by both modes will be consistent with Figs.~\ref{fig:Simulation}(k) and \ref{fig:Simulation}(u). From the results, it can be concluded that our method is suitable for both gray-scale and color scenes.

To obtain a quantitative measure of the image quality, here we introduce the peak signal-to-noise ratio (PSNR) as a figure of merit, which is defined as
\begin{equation}
\textrm{PSNR}=10\log(255^2/\textrm{MSE}),
\label{eq:PSNR}
\end{equation}
where $\textrm{MSE}=\frac{1}{pq}\sum\nolimits_{i,j=1}^{p,q}[\tilde U(i,j)-U_o(i,j)]^2$ describes the squared distance between the recovered image and the original image, $U_o$ and $\tilde U$ stands for the original image and the reconstructed image, respectively, all of $p\times q$ pixels. Naturally, the larger the PSNR value, the better the quality of the image recovered. According to the definition of PSNR, all the recovered images should be normalized to a range of $0\sim255$.

\begin{figure}[htbp]
\centering\includegraphics[width=0.95\linewidth]{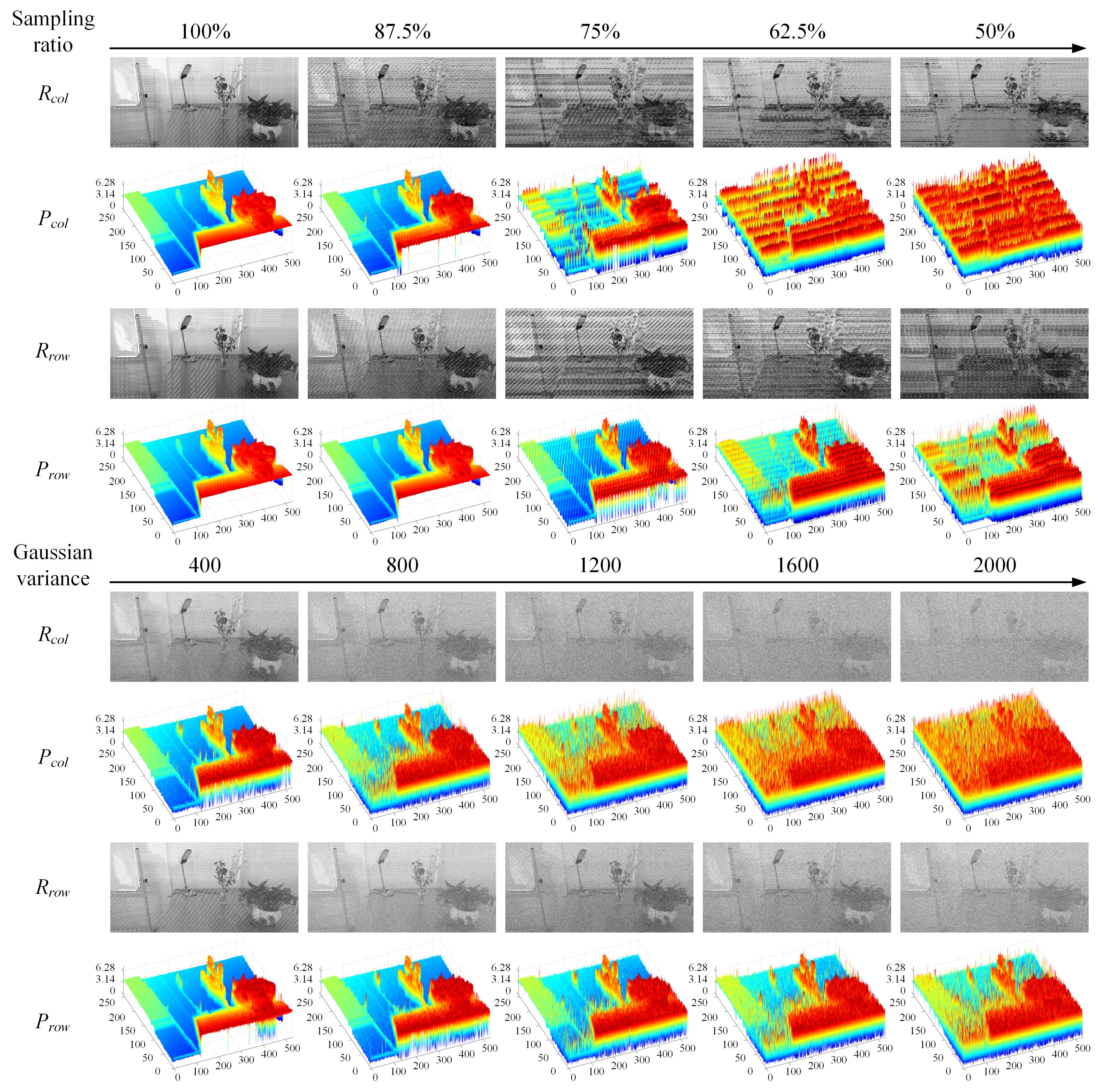}
\caption{Simulated results of our method with compressive sampling (see the top four rows) and the reconstructed results of the Hadamard full sampling under different variances of additive Gaussian noise (see the bottom four rows). The sizes of all images are $256\times532$. The sampling ratio is changing from 100\% to 50\% with a 12.5\% stepping decrease. The Gaussian variance is set to 400, 800, 1200, 1600 and 2000, while the mean of the whole measured values is 128.5 for both mode-col and mode-row. $R_{col}$ and $P_{col}$ stand for the recovered surface reflectance images and phase angle 3D maps of mode-col, while $R_{row}$ and $P_{row}$ denote the computed surface reflectance images and phase angle 3D maps of mode-row.}
\label{fig:Analysis}
\end{figure}

For compressive sampling, we also perform some simulation under different sampling ratios, which range from 100\% to 50\% (the limit) with a 12.5\% stepping. Here the sampling rate is defined as the ratio of the number of measurements to the number of pixels. The $R_{col}$ and $P_{col}$ results in mode-col and mode-row can be found in the top four rows of Fig.~\ref{fig:Analysis}. It is not difficult to find from the figures that as the number of measurements decreases, the quality of both $R_{col}$ and $P_{col}$ reconstructions is gradually degraded.

Then the Gaussian noise with different variances is added to the measured values $y$. For better quantitative analysis, here we use Hadamard full sampling rather than compressive sampling. The recovered results of $R_{col}$ and $P_{col}$ in mode-col and mode-row are given in the bottom four rows of Fig.~\ref{fig:Analysis}. Especially the result of phase recovery is a good example demonstrating its robustness to noise.

\begin{figure}[htbp]
\centering\includegraphics[width=0.95\linewidth]{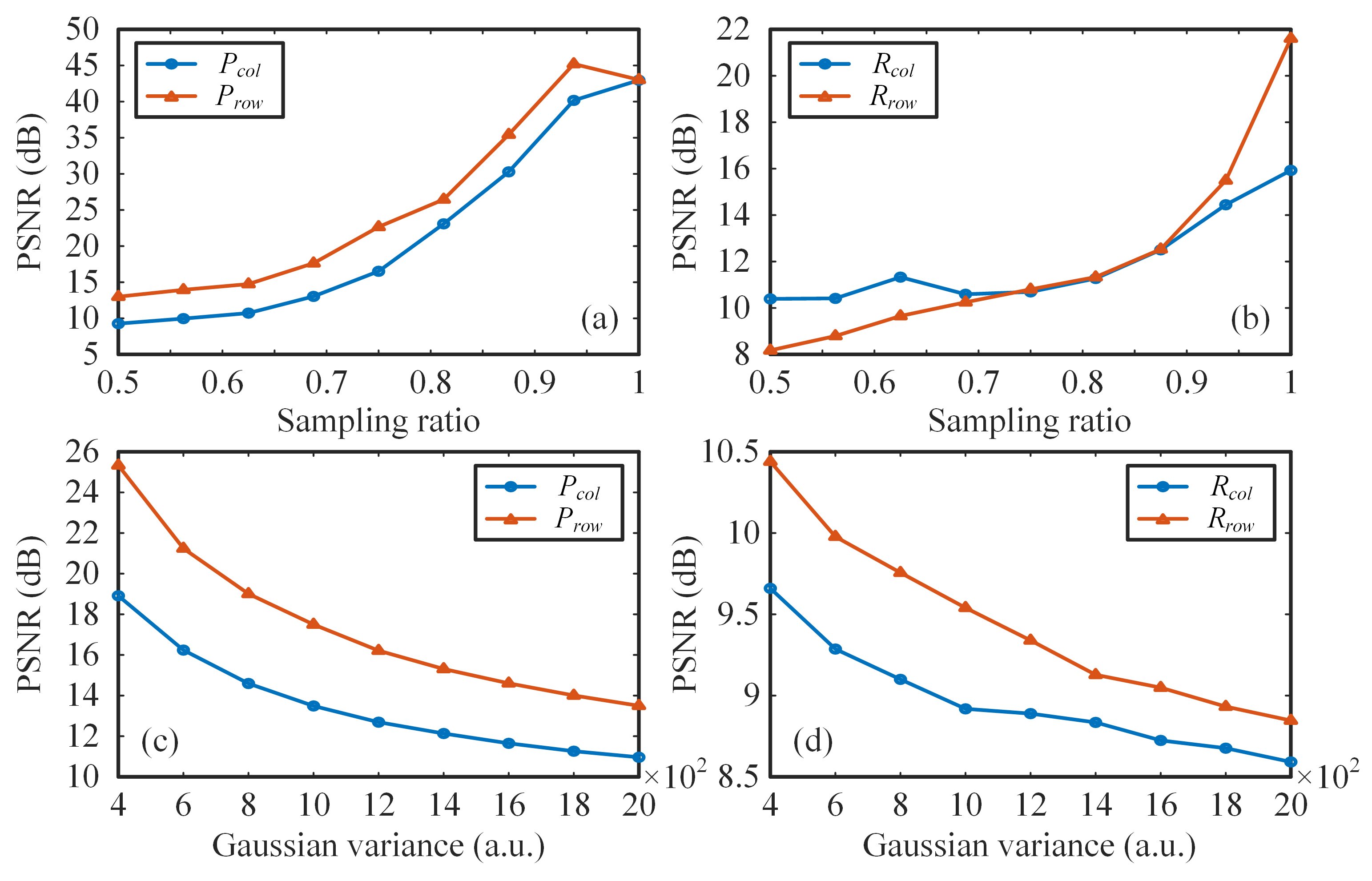}
\caption{PSNR curves of our method vs. (a)--(b) the sampling ratio and (c)--(d) the variance of the additive Gaussian noise. In each graph, we have compared the data of both mode-col and mode-row. (a) and (c) are the PSNR curves of recovered phase angle maps, while (b) and (d) are the PSNR curves of retrieved surface reflectance images.}
\label{fig:PSNR}
\end{figure}

We have drawn the PSNR curves of above two simulated results, as depicted in Fig.~\ref{fig:PSNR}. The PSNR values as a function of the sampling ratio is presented in Figs.~\ref{fig:PSNR}(a)--(b) while the ones as a function of the Gaussian noise is given in Figs.~\ref{fig:PSNR}(c)--(d). From the graph, it is clear that the image quality of mode-row is better than that of mode-col with a high probability, because the signal length of each scanning column in the mode-row is four times the one in the mode-col. The image quality is proportional to the sampling rate and inversely proportional to the Gaussian noise variance. Besides, it is interesting to find that the larger the number of vertical pixels of the scene, the better the reconstruction quality.

\section*{\textsf{Discussion and Proposal for Experimental Setup}}
We have considered merging the two sets of patterns, which means that if you look at the top view facing the motion direction, a long-period jigsaw-puzzle-reorganized sinusoidal strip pattern is floating over a time-invariant measurement pattern with its line stripe loop moving down. Here the aforementioned period refers to the cyclic repetition period of the entire sinusoidal stripe pattern, rather than the sinusoidal fringe spatial period itself. The speed of the sinusoidal pattern translation is exactly the same as the moving speed of the scene, that is, $1/4$ or 1 pixel moving forward for each measurement. As for its realization in a real optical system, we strongly suggest using the most popular programmable digital micromirror device (DMD), consisting of millions of micromirrors (pixels), each of size 13.68~$\mu$m$\times$13.68~$\mu$m. Each mirror is orientated either $12^\circ$ (bright pixel 1) or $-12^\circ$ (dark pixel 0) with respect to the normal of the DMD plane, determined by a preloaded sequence of modulated binary patterns. It is known that the  DMD is the fastest spatial light modulator for the time being, with a modulation rate up to 32,550~Hz (patterns/s). The matrix loaded onto the DMD consists of either 0 or 1 while the sinusoidal pattern is generally with 8-bit gray-scale. In order to get the bit depth on the DMD, we can either use pulse width modulation (PWM) or motion blur method \cite{Jiang2017} which dithers binary coded triangular patterns along the slope direction of sinusoidal stripes. The reflected light from the DMD can be treated as computational illumination light to be projected onto the moving scene, and the retroreflected light will be sampled by a line array detector via a cylindrical lens. Additionally, the deterministic pattern used in our method is composed by $\pm1$, it can be realized by subtle shifting and stretching the matrix $T$ (which can be either $A$ or $H$) into two complementary matrices $\hat{T}=(1+T)/2$ and $\check{T}=(1-T)/2$, whose difference matrix is exactly $T$. For PWM method, we can modulate one gray-scale pattern $\hat{T}_{gray}$ immediately followed by its inverse (complementary) gray-scale pattern $255-\hat{T}_{gray}$. For motion blur method, we can display one binary pattern then its complementary one. The corresponding measured values also need to be differentiated. We named this approach ``positive-negative" intensity modulation \cite{YuSR2014,YuOC2016}, as it allows one to compensate for the mean shift of the sampled signal and to generate a zero-mean sensing matrix. As mentioned before, the system can either use two projectors scheme or only one projector scheme, the results will be the same. The constructions of other arrangement matrices and the combination with other phase-shifting methods will be our next work. We have reasons to believe that the experimental realization of our system in the future will fill in the technical gap and promote the practical development of SPI in the 3D fields, like baggage screening on the conveyor belt, product inspection on the assembly line, 3D reproduction of artifacts, virtual vision, autonomous driving, and so on.

\section*{\textsf{Conclusion}}
In conclusion, this paper introduces a 3D imaging method based on deterministic measurements of the relative moving scene, which is projected with a single-frame jigsaw-puzzle-reorganized sinusoidal pattern and a measurement pattern simultaneously, without the need of repeating the scene motion multiple times. The sinusoidal pattern translates with the scene, while the deterministic measurement pattern is time-invariant. Our scheme is capable of achieving parallel SPI by placing a linear array detector along (instead of across) the motion axis of the scene and performing multi-pixel axial flat brush scanning. Therefore, the technique takes full advantage of spatial multiplexing of measurements in which a pixel does not need to stare at the same part of the scene and sample it a plurality of times, getting rid of motion blur problem. By using distributed parallel computing, we can recover all the column signals of the scene in parallel, which brings possibilities for real-time 3D imaging. The wrapped phase map of the desired scene can be acquired according to the jigsaw-puzzle-reorganized matrices, with the use of four-step phase shifting technique, which offers noise robustness. Then unwrapping the phase can produce the phase angle map together with its 3D display and the surface reflectivity. The deterministic measurement pattern can either use complete Hadamard matrix (full sampling) or a cake-cutting sort of Hadamard basis (compressive sampling with less pixels), accelerating matrix calculation, and the latter also minimizes redundancy caused by traditional random sampling. The proposal of using a DMD for complex illumination may bring feasibilities into practical applications, including 3D item detection on conveyor belts, airborne 3D navigation, 3D reconstruction for autonomous drive, 3D reproduction of artifacts, and the like.

\section*{\textsf{Acknowledgments}}
This work is supported by the National Natural Science Foundation of China (61801022), the Beijing Natural Science Foundation (4184098), the National Key Research and Development Program of China (2016YFE0131500), the International Science and Technology Cooperation Special Project of Beijing Institute of Technology (GZ2018185101), and the Beijing Excellent Talents Cultivation Project - Youth Backbone Individual Project. The author warmly appreciate Shuo-Fei Wang and Xiao-Peng Jin for useful discussions.

\section*{\textsf{Author Contributions}}
W.-K.Y. conceived the idea, designed the algorithms, performed the simulations, analyzed the data, and wrote this manuscript.

\end{document}